\documentclass[a4paper,twocolumn,english,showpacs,preprintnumbers,amsmath,amssymb,aps,prl,notitlepage,superscriptaddress]{revtex4-1}
\usepackage[latin9]{inputenc}
\usepackage{mathtools}
\usepackage{amsmath}
\usepackage{mathdots}
\usepackage{graphicx}

\makeatletter

\pdfpageheight\paperheight
\pdfpagewidth\paperwidth

\DeclareTextSymbolDefault{\textquotedbl}{T1}

\usepackage[english]{babel}
\usepackage{soul}
\usepackage{cleveref}
\usepackage{bbold}
\usepackage{multirow}
\usepackage{colortbl}
\definecolor{kugray5}{RGB}{224,224,224}

\usepackage{dcolumn}
\usepackage{bm}

\usepackage{color}
\usepackage{amstext}
\usepackage{braket}
\usepackage{tikz}
\usepackage{physics}
\usepackage{pdfsync}
\usetikzlibrary{matrix,decorations.pathreplacing}

\makeatother

\usepackage{babel}
\begin{document}
\title{1D topological insulators with non-centered inversion symmetry axis}
\author{A. M. Marques}
\email{anselmomagalhaes@ua.pt}
\affiliation{Department of Physics $\&$ I3N, University of Aveiro, 3810-193 Aveiro,
Portugal}

\author{R. G. Dias}
\email{rdias@ua.pt}
\affiliation{Department of Physics $\&$ I3N, University of Aveiro, 3810-193 Aveiro,
Portugal}
\date{\today}
\begin{abstract}
In this paper, we discuss the characteristic features 
of 1D topological insulators with inversion symmetry but non-centered
inversion axis in the unit cell, for any choice of the unit cell.
In these systems, the global inversion operation generates
a $k$-dependent inversion operator within the unit
cell and this implies a non-quantized Zak's phase both for the trivial
and non-trivial topological phases. By relating the Zak's phase  with the eigenvalues of modified parity operators  at the inversion invariant momenta, a corrected quantized form of the Zak's phase is derived.
We show that finite energy topological edge states of this family of chains are symmetry-protected not by usual chiral symmetry but by a hidden sublattice chiral-like symmetry. A simple justification is presented for shifts in the polarization quantization relation for any choice of endings of these chains.
\end{abstract}

\maketitle

In one-dimensional (1D)  topological insulators with  unit cells that respect inversion($\mathcal{I}$)-symmetry, the eigenstates of the Bloch Hamiltonian $H_k$  generate symmetric charge distribution in relation to the unit cell inversion center. Choosing open boundary conditions (OBC) commensurate with the unit cell,  polarization is due only to the edge states contribution and  the bulk-edge correspondence can be described by the intercell Zak's phase that ignores the relative position of orbitals within the same unit cell.
In order to find  topological invariants that  protect finite energy edge states in the case of non-commensurate OBC or non-centered
$\mathcal{I}$-axis in the unit cell, modified approaches have been proposed  such as the splitting of the Zak's phase into intracell and intercell contributions \cite{Springborg2004,Kudin2007,Miert2017,Rhim2017,Lin2018}, the squaring of the Hamiltonian \cite{Arkinstall2017,Kremer2018,Midya2018,Zhang2019,Pelegri2019,Pelegri2019b} or synthetic dimensions \cite{Mei2012,Lang2012,Zhu2013,Qin2017,Alvarez2019}.

In this paper, we propose a different path to address 1D topological insulators with $\mathcal{I}$-symmetry but non-centered
$\mathcal{I}$-axis in the unit cell, for any choice of the unit cell. 
Relying in the Wilson's loop method \cite{Alexandradinata2014,Asboth2016}, a  quantized corrected Zak's phase is related with the eigenvalues of modified parity operators  at the $\mathcal{I}$-invariant momenta. 
Finite energy edge states in these models are shown to be eigenstates of (and therefore protected by) a $\hat{C}_{1/2}$ operator that reflects a underlying sublattice chiral symmetry.
Furthermore, the non-centered $\mathcal{I}$-axis in the unit cell implies a displacement of the inversion center of the bulk charge distribution  with relation to the center of the chain for commensurate OBC generating a shift in the polarization quantization.

\emph{Generalization of Zak's phase}: 
Given a 1D tight-binding Hamiltonian with translation invariance, its eigenstates are Bloch states  $\vert k \rangle \otimes \vert u_k \rangle$, where $k$ is the momentum and $\vert u_k \rangle$ is the respective eigenstate of the Bloch Hamiltonian $H_k$. A convenient way to compute the Zak's phase in band $j$ is through
the Wilson loop, $\mathcal{W}_{j}=\prod_{n=0}^{N-1}\braket{u_{j}(-\pi+n\Delta k)}{u_{j}(-\pi+(n+1)\Delta k)},$
where we have set the momentum increment to $\Delta k=\frac{2\pi}{N_{uc}}$,
with $N_{uc}$ the number of unit cells in the periodic chain. In the continuous
limit ($N_{uc}\to\infty$, $\Delta k\to0$), the Zak's phase \cite{Zak1989} of band $j$
becomes 
\begin{equation}
\gamma_{j}=i\int_{-\pi}^{\pi}dk\bra{u_{j}(k)}d_{k}\ket{u_{j}(k)}=\mbox{Arg}\big(\lim_{N\to\infty}\mathcal{W}_{j}\big)\label{eq:zakarg}
\end{equation}
where $d_{k}$ is the $k$ derivative.

\begin{figure}[t]
\begin{centering}
\includegraphics[width=0.5\textwidth]{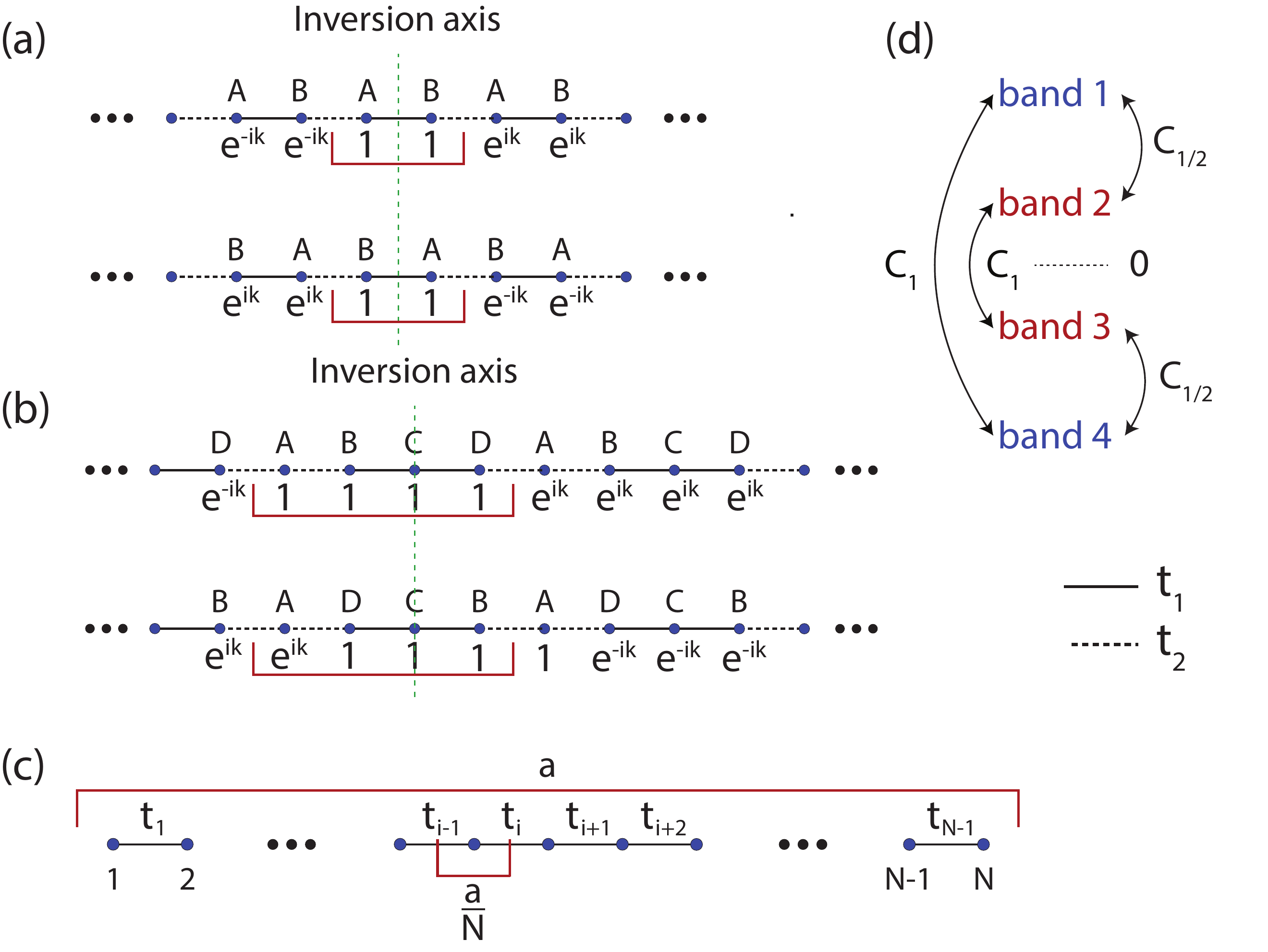} 
\par\end{centering}
\caption{(a) Scheme of an inversion operation in a periodic SSH ($t_{1}t_{2}$)
chain. Both sites in a given unit cell have the same phase, before
and after the inversion. (b) Scheme of an inversion operation in a
periodic $t_{1}t_{1}t_{2}t_{2}$ chain. After the inversion is performed,
the A site in each unit cell gains an extra phase factor of $e^{ik}$,
relative to the other sites, as a consequence of the $\mathcal{I}$-axis
being to the right of the center. For this model, no choice of unit
cell has a centered $\mathcal{I}$-axis. (c) Unit cell of arbitrary size $a$ and hoppings configuration. All $N$ sites are uniformly spaced, with an intersite spacing of $\frac{a}{N}$. (d) Action of the unitary Hermitian operators
$\hat{C}_{1}$ and $\hat{C}_{1/2}$ in the band structure of the  $t_{1}t_{1}t_{2}t_{2}$ chain. }
\label{fig:invsym}
\end{figure}

Eigenstates with opposite momenta are related by the $\mathcal{I}$-operator
within the unit cell as $\ket{u_{j}(-k)}=e^{i\theta_{k}}\hat{\pi}_{k}\ket{u_{j}(k)}$,
for all $k\neq0,\pi$, where $e^{i\theta_{k}}$ is an arbitrary phase
factor that we take out for convenience from now on since they will
appear as conjugate pairs in the Wilson's loop \cite{Asboth2016}. The
condition for a $k$-independent $\hat{\pi}$ is that the $\mathcal{I}$-axis is at the center of the considered unit cell [see Fig.~\ref{fig:invsym}(a)]. There are models,
however, where this condition is not met for any choice of the unit
cell, as demonstrated in the case of Fig.~\ref{fig:invsym}(b). In
the case of the arbitrary unit cell of Fig.~\ref{fig:invsym}(c),
both in size and morphology (regarding the hoppings parameters), of
uniformly spaced sites [at positions $r_{j}=(j-\frac{1}{2})\frac{a}{N_a}$,
with $j=1,2,...,N_a$], the possible positions for the $\mathcal{I}$-axis
are given by $r_{m}=a(\frac{1}{2}+\frac{m}{2N_a})$, with $m=0,\pm1,...,\pm N_a$
indicating its displacement from the center of the unit cell. A $k$-dependence
in the $\mathcal{I}$-operator appears for $m\neq0$, with a general
form in the ${\vert k,j\rangle}$ basis (where $j$
labels the sites in the unit cell) being given by 

\begin{equation}
\begin{split}\hat{\pi}_{k} & =\Theta(m)\left[\begin{array}{c|c}
e^{ik}\left[\hat{\pi}\right]^{m\times m} & 0\\
\hline 0 & \left[\hat{\pi}\right]^{(N_a-m)\times(N_a-m)}
\end{array}\right]\\
 & +\Theta(-m)\left[\begin{array}{c|c}
\left[\hat{\pi}\right]^{(N_a+m)\times(N_a+m)} & 0\\
\hline 0 & e^{-ik}\left[\hat{\pi}\right]^{\vert m\vert\times\vert m\vert}
\end{array}\right],
\end{split}
\end{equation}
where $\left[\hat{\pi}\right]^{m\times m}$ is the usual matrix representation
of the $\mathcal{I}$-operator in a unit cell with $m$ sites, that is,
a skew diagonal matrix of dimension $m$ such that $\left[\hat{\pi}\right]_{ij}^{m\times m}=\delta_{i,m+1-j}$.

The model in Fig.~\ref{fig:invsym}(b) corresponds to the particular
case of $N_a=4$ and $m=1$. Assuming this case (general
expressions will be given below), and keeping only the first order
terms in the Wilson's loop, we obtain
\begin{eqnarray}
\braket{u_{j}(k-dk)}{u_{j}(k)} & = & 1+i\delta\phi,\label{eq:ukpos}\\
\braket{u_{j}(-k)}{u_{j}(-k+dk)}%
 & \simeq & 1-i\big(\delta\phi+dk\vert u_{j,A}(k)\vert^{2}\big),
\end{eqnarray}
for $dk<k<\pi$. In the last step we assumed, to leading order, $\braket{u_{j,A}(k)}{u_{j,A}(k-dk)}\approx\vert u_{j,A}(k)\vert^{2}$.
However, different relations hold at the $\mathcal{I}$-invariant
momenta $k=0,\pi$, where $\ket{u_{j}(-\pi)}\equiv\ket{u_{j}(\pi)}$,
\begin{eqnarray}
\braket{u_{j}(-dk)}{u_{j}(0)}%
 & \simeq & \bra{u_{j}(dk)}\hat{\pi}_{0}^{\dagger}\ket{u_{j}(0)}\nonumber \\
 & - & idk\braket{u_{j,A}(dk)}{u_{j,A}(0)},\\
\braket{u_{j}(-\pi)}{u_{j}(-\pi+dk)}%
 & \simeq & \bra{u_{j}(dk)}\hat{\pi}_{\pi}\ket{u_{j}(\pi-dk)}\nonumber \\
 & + & idk\braket{u_{j,A}(\pi)}{u_{j,A}(\pi-dk)}.
 \label{eq:ukpi}
\end{eqnarray}
The modified parity of the corresponding eigenstates is well defined,
that is, $\hat{\pi}_{0}\ket{u_{j}(0)}=P_{0}\ket{u_{j}(0)}$ and $\hat{\pi}_{\pi}\ket{u_{j}(\pi)}=P_{\pi}\ket{u_{j}(\pi)}$,
with $P_{0},P_{\pi}=\pm1$. Note that  $\hat{\pi}_{0}$ and $\hat{\pi}_{\pi}$ are different and modified parity operators. The procedure now is to substitute (\ref{eq:ukpos}-\ref{eq:ukpi})
in the computation of the Wilson loop in (\ref{eq:zakarg}) to obtain
the following simplified expression for the Zak's phase, 
\begin{equation}
\gamma_{j}=\mbox{Arg}(P_{0}P_{\pi})-\int_{0}^{\pi}dk\vert u_{j,A}(k)\vert^{2},\label{eq:zakphase}
\end{equation}
which is in general \textit{non-quantized} due to the last term. The
last term in (\ref{eq:ukpi}), with a positive sign, was disregarded
as an infinitesimal surface term. 
A $\pi$-quantized Zak's phase in each band, $\tilde{\gamma}_{j}$,
can still be recovered by dropping the last term in the previous equation, 
so the general expression for $\tilde{\gamma}_{j}$
is 
\begin{eqnarray}
\gamma_{j} & = & i\int_{-\pi}^{\pi}dk\bra{u_{j}(k)}d_{k}\ket{u_{j}(k)},\\
\tilde{\gamma}_{j} & = & \begin{cases}
\gamma_{j}+\sum_{s=1}^{m}\int_{0}^{\pi}dk\vert u_{j,s}(k)\vert^{2},\text{\ \ \ \ \ \ \ \ for\ }m>0,\\
\\
\gamma_{j}-\sum_{s=0}^{\vert m\vert-1}\int_{0}^{\pi}dk\vert u_{j,N-s}(k)\vert^{2},\text{\ for\ }m<0.
\end{cases},\label{eq:generalmodzak}
\end{eqnarray}
which agrees with $\tilde{\gamma}_{j}=\mbox{Arg}(P_{0}P_{\pi}),$
for all $m$. 

\emph{$C_{1}$ chiral symmetry and $C_{1/2}$ symmetry}: Again, we
will focus on a specific example with non-centered $\mathcal{I}$-axis:
the $t_{1}t_{1}t_{2}t_{2}$ model, which is a particular case of the
SSH$_{4}$ model \cite{Eliashvili2017,Maffei2018}. Considering the unit cell of
Fig.~\ref{fig:invsym}(b), the bulk Hamiltonian (intercell spacing
was set to $a=1$) is 
\begin{equation}
H_{Bloch}(k)=\begin{pmatrix}0 & t_{2} & 0 & t_{2}e^{-ik}\\
t_{2} & 0 & t_{1} & 0\\
0 & t_{1} & 0 & t_{1}\\
t_{2}e^{ik} & 0 & t_{1} & 0
\end{pmatrix}.\label{eq:hamiltt1t1t2t2}
\end{equation}
The eigenvalue equation, $H_{Bloch}(k)\ket{u_{j}(k)}=E_{j}(k)\ket{u_{j}(k)}$,
yields four bands, $E_{1}(k)=-E_{4}(k)=\sqrt{\Delta_{0}+\vert\Delta_{k}\vert}$
and $E_{2}(k)=-E_{3}(k)=\sqrt{\Delta_{0}-\vert\Delta_{k}\vert}$,
with $\Delta_{k}=t_{1}^{2}+t_{2}^{2}e^{ik}=\vert\Delta_{k}\vert e^{i\phi_{k}}$
and $\cot\phi_{k}=\frac{t_{1}^{2}}{t_{2}^{2}\sin k}+\cot k.$ Note
that $\pm\vert\Delta_{k}\vert$ is the energy dispersion of the SSH
model with staggered squared hoppings $t_1^2$ and $t_2^2$. 
In the following, we set $t_{2}=1$ as the energy unit and $t_{1}/t_{2}=t$.

After some algebra, the top band eigenvector can be cast as 
\begin{eqnarray}
\ket{u_{1}(k)} & = & \frac{1}{2}\begin{pmatrix}\sqrt{2}\sin\theta_{k}\\
e^{i(k-\phi_{k})/2}\\
\sqrt{2}\cos\theta_{k}e^{ik/2}\\
e^{i(k+\phi_{k})/2}
\end{pmatrix},\label{eq:u1}
\end{eqnarray}
with 
\begin{equation}
\theta_{k}=\arctan(\frac{1-t^{2}+\sqrt{1+t^{4}+2t^{2}\cos k}}{t\sqrt{2(1+\cos k)}}),
\end{equation}
and $k,\phi_{k},\theta_{k}\in]-\pi,\pi]$. The other three eigenvectors
are obtained from this one using  two unitary Hermitian operators
$\hat{C}_{1}=\sum\limits _{k}\hat{C}_{1}(k)$ and $\hat{C}_{1/2}=\sum\limits _{k}\hat{C}_{1/2}(k)$,
with
\begin{eqnarray*}
\hat{C}_{1}(k) & = & \ket{u_{1}(k)}\bra{u_{4}(k)}+\ket{u_{2}(k)}\bra{u_{3}(k)}+H.c.,\\
\hat{C}_{1/2}(k) & = & \ket{u_{1}(k)}\bra{u_{2}(k)}+\ket{u_{3}(k)}\bra{u_{4}(k)}+H.c.,
\end{eqnarray*}
which in the basis ${\vert k,A\rangle,\vert k,C\rangle,\vert k,B\rangle,\vert k,D\rangle}$
(note the reordering) are given by $\hat{C}_{1}(k)=\sigma_{z}\otimes\mathbb{\mathbb{1}}_{2}$
and 
\begin{equation}
\hat{C}_{1/2}(k)=\left[\begin{array}{cc}
\text{sgn}(k) [\sin(\frac{k}{2})\sigma_{x}-\cos(\frac{k}{2})\sigma_{y} ]& 0\\
0 & \sigma_{z}
\end{array}\right],
\end{equation}
where $\sigma_{\alpha}$ are the Pauli matrices. In all eigenstates,
the occupation probability in both sublattices AC and BD is 1/2.

The operators $\hat{C}_{1}$ and $\hat{C}_{1/2}$, whose action is depicted in Fig.~\ref{fig:invsym}(d), reflect respectively
the usual chiral symmetry and a hidden sublattice chiral-like symmetry of this
system. All bipartite models have chiral symmetry, defined by the
existence of an operator $\hat{C}_{1}$ such that $\hat{C}_{1}H\hat{C}_{1}=-H$
(note that this condition does not define $\hat{C}_{1}$ uniquely).
The presence of chiral symmetry entails a symmetric energy spectrum
around zero. In particular, if a chiral-symmetric model has edge
states with non-zero energy, as we will show to be the case in the
$t_{1}t_{1}t_{2}t_{2}$ model, these edge states appear in chiral
pairs with symmetric energies and are localized at the same edge \cite{Ryu2002}.
Chiral symmetry, by itself, only ensures the topological protection
of edge states if these have zero energy, otherwise the topological
protection of the edge states has to be defined with recourse to another
operator.

Let us determine the general form of an edge-like eigenstate of an infinite
$t_{1}t_{1}t_{2}t_{2}$ chain. These states are in general non-normalizable.
However, if they have zeros of amplitude at certain sites, we can
cut the infinite chain at these sites and these edge states become
exact eigenstates of the chain with OBC, orthogonal to the harmonic
bulk states, since the boundary conditions are automatically satisfied.
A general eigenstate of the infinite $t_{1}t_{1}t_{2}t_{2}$ chain
can be written as $\ket{w(\varepsilon,c)}=\sum_{j}c^{j}\ket{u(\varepsilon,c)}$, where
$c=e^{ik},k\in\mathbb{C}$, that is, we allow for a complex momentum
$k$ so that exponentially decaying solutions (the edge states) are
not ruled out \cite{Delplace2011,Banchi2013,Hugel2014,Duncan2018}. 
The eigenvalue equation $H\ket{w(\varepsilon,c)}=\varepsilon\ket{w(\varepsilon,c)}$
can be rewritten as $H_{edge}(c)\ket{u(\varepsilon,c)}=\varepsilon\ket{u(\varepsilon,c)}$
with a non-Hermitian $H_{edge}(c)=H_{Bloch}(e^{ik}\rightarrow c)$
which yields the eigenenergies $\varepsilon=\varepsilon_{k}\left(e^{ik}\rightarrow c\right)$
where $\varepsilon_{k}$ is the energy dispersion of the Bloch states
given in the previous section, written as $\varepsilon_{k}=\pm\sqrt{\Delta_{0}\pm\sqrt{\Delta_{k}\Delta_{k}^{*}}}$
(the four sign combinations are possible).
The respective (non-normalized) eigenstates are
\begin{equation}
\ket{u(\varepsilon,c)}=\begin{pmatrix}u_{A}(\varepsilon,c)\\
u_{B}(\varepsilon,c)\\
u_{C}(\varepsilon,c)\\
u_{D}(\varepsilon,c)
\end{pmatrix}=\begin{pmatrix}\varepsilon(\varepsilon^{2}-2t^{2}))\\
\varepsilon^{2}-t^{2}+ct^{2}\\
(1+c)t\varepsilon\\
c\varepsilon^{2}+t^{2}-ct^{2}
\end{pmatrix}.\label{eq:eigenedge}
\end{equation}
We search for the values of $c$ for which (\ref{eq:eigenedge}) has
a zero of amplitude on at least one of the components in any of the four possible eigenstates, as required by the OBC. There are
only four such values of $c$: (i) $c=\pm1$ corresponding to the
$k=0$ and $k=\pi$ bulk states of the four bands; (ii)$\ket{u(\varepsilon,c=-\frac{1}{t^{2}})}=(-t^{2},0,t,\varepsilon)^{T}$
and $\ket{u(\varepsilon,c=-t^{2})}=(1,\varepsilon,t,0)^{T}$ with
energy $\varepsilon=\sqrt{t^{2}+1}$, obtained by algebraic development of \eqref{eq:eigenedge} \footnote{See Supplemental Material for additional details on deriving the edge states from the general form of (\ref{eq:eigenedge})}, as well as the respective chiral pairs.
The available edge states of this model are given by the 
$c=-t^{2},-\frac{1}{t^{2}}$ cases, with $t\neq1$,
each with two possible edge states with symmetric energies.

In the SSH model, for $t<1$, the left and right edge states are written as $\ket{L} \approx \sum_j (-t)^{j-1} \sqrt{1-t^2} \ket{j,A}$
and $\ket{R}\approx\sum_j (-1/t)^{j-1} \sqrt{1-1/t^2} \ket{j,B}$. These states are their own chiral
pairs, that is, $C_{1}\ket{L(R)}=\sigma_{z}\ket{L(R)}=+(-)\ket{L(R)}$
and it is this property that forces the energy of these states to
remain zero as long as the chiral symmetry is not broken. The finite
energy of edge states of the $t_{1}t_{1}t_{2}t_{2}$ chain described
above implies that these states are not protected by the usual chiral
symmetry, but they are by the symmetry associated with the $\hat{C}_{1/2}$
operator. An important detail is that, unlike $\hat{C}_{1}$, the
$\hat{C}_{1/2}$ operator has a $k$-dependent $\hat{C}_{1/2}(k)$
matrix representation. In the edge states subspace, the matrix representation
of the $\hat{C}_{1/2}$ operator is obtained in the same way as $H_{edge}$,
that is, $\hat{C}_{1/2}^{edge}(c)=\hat{C}_{1/2}^{Bloch}(e^{ik}\rightarrow c)$ \footnote{See Supplemental Material for additional details on the derivation of $\hat{C}_{1/2}^{edge}(c)$}
and in the $\{\ket{A},\ket{C},\ket{B},\ket{D}\}$ basis of each unit cell, it is given
by
\begin{equation}
\hat{C}_{1/2}^{edge}(c)=\begin{bmatrix}0 & \frac{1}{\sqrt{\vert c\vert}} & 0 & 0\\
\sqrt{\vert c\vert} & 0 & 0 & 0\\
0 & 0 & 1 & 0\\
0 & 0 & 0 & -1
\end{bmatrix}.
\end{equation}
 \emph{The finite energy edge states of the $t_{1}t_{1}t_{2}t_{2}$ chain are
their own pairing state under $\hat{C}_{1/2}^{edge}(c)$}.

If we square the Hamiltonian of the OBC $t_{1}t_{1}t_{2}t_{2}$ chain shown in Fig.~\ref{fig:invsym}(b) (which has $\mathcal{I}$-symmetry but non-commensurate OBC), one obtains two decoupled OBC chains with $\mathcal{I}$-symmetry: a commensurate SSH chain  (with B and D sites) with hopping parameters $t_1^2$ and $t_2^2$  and local potential $t_1^2+ t_2^2$ and a chain  (with A and C sites)  with non-commensurate OBC,  staggered local potentials, and an impurity potential at the ends. Both chains have the same spectrum and the edge state in the SSH chain is protected by the usual chiral symmetry (which is a sublattice symmetry in the original chain) if  zero energy is set  at the value of  the diagonal local potential. This SSH chiral symmetry corresponds to the lower block in the previous matrix representation of $\hat{C}_{1/2}^{edge}(c)$, and accordingly the edge states of the $t_{1}t_{1}t_{2}t_{2}$ chain are protected against perturbations that preserve this lower $\sigma_z$ block, such as hoppings terms connecting B and D sites.
Note that such perturbations break the chiral symmetry given by the $\hat{C}_{1}$ operator, showing unambiguously that the chiral-like symmetry given by the $\hat{C}_{1/2}^{edge}(c)$ operator is the topologically protecting symmetry of our model. 

\begin{figure}
\includegraphics[width=.45\textwidth]{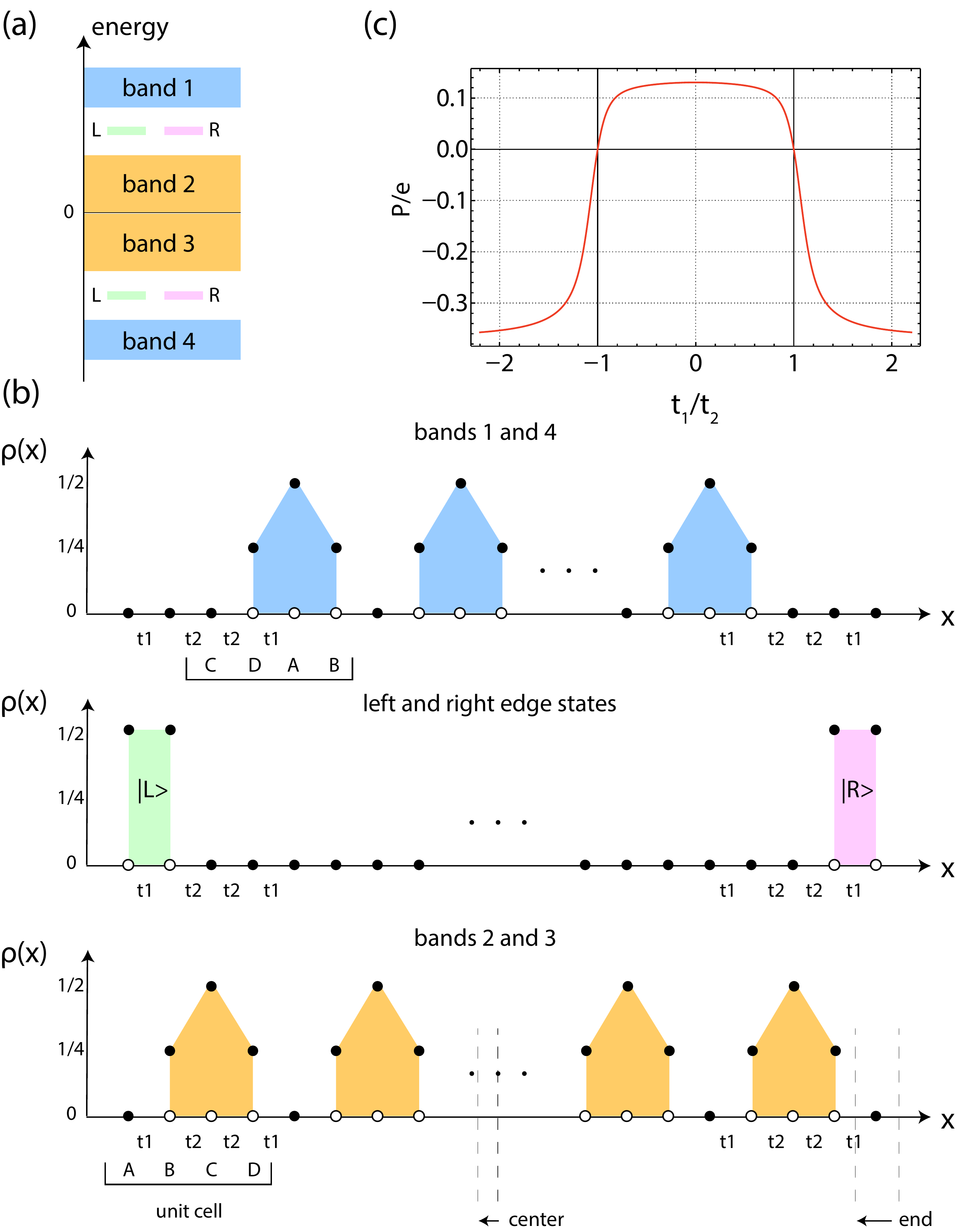}
\caption{
(a) Band structure and (b) charge distribution  (in units of the electron charge) of a $t_1t_2t_2t_1$ chain (for $t_1\gg t_2$) with $\mathcal{I}$-symmetry. In this case, the number of unit cells is non-integer, since an extra A site was added at the right end. The colors indicate which bands/edge states generate the charge distributions shown at the bottom. The triangular blocks reflect the nearly compact Wannier states of the bulk bands. In order to address  $t_1\ll t_2$, the colors of the bands should be exchanged and the edge states removed from the  band gaps and moved into the bulk of bands 2 and 3. (c) Polarization per unit length for the  $t_1t_2t_2t_1$ chain in the case of 6 unit cells, showing  the shift in the quantization relation. \label{fig:cartoon}}
\end{figure}

\emph{Non-quantized polarization}: If the unit cell is commensurate with the OBC, real-space $\mathcal{I}$-symmetry is absent in 1D topological insulators with non-centered
$\mathcal{I}$-symmetry axis within the unit cell. $\mathcal{I}$-symmetry in relation to the center of the chain can be recovered considering  a non-integer number of unit cells [for example, adding an extra A site at the right end of the $t_1t_2t_2t_1$ chain, see  Fig.~\ref{fig:cartoon}(b)].
At the topological transition point of the   $t_1t_2t_2t_1$ chain, $t_1=t_2$, both cases show zero polarization when any number of bands/states are occupied. 

Assuming the latter case and $t_1\gg t_2$, one has left and right finite energy edge states in the top and bottom gaps as shown in Fig.~\ref{fig:cartoon}(a) and the charge distribution generated by the bands and the edge states has approximately the simple form displayed on Fig.~\ref{fig:cartoon}(b) (the colors give the correspondence to the respective bands/edge states). 
The polarization when  only the bottom band and an edge state are occupied is $\pm e/2$ depending on whether the left or the right edge state is occupied. This is expected since $\mathcal{I}$-symmetry is present. The former case (with integer number of unit cells) is obtained from the latter dropping the A site at the right end of the chain generating a shift of the center of the chain as shown in the bottom of Fig.~\ref{fig:cartoon}(b). The charge distribution due to the bottom band is not significantly  affected when removing that  site, but the right edge state is,  becoming a zero energy edge state nearly fully localized at the D site at the right end. 

Furthermore, the inversion center of the bottom band charge distribution becomes shifted in relation to the center of the chain, contributing therefore to the polarization. This contribution can be determined in two ways, either by summing charge times position throughout the chain (in this case the contribution of the positive background is zero) or by adding the polarization of the unit cells and in this case the contribution of the positive background at the empty sites at the ends of the chain must be added if one does not have the same number of empty sites outside the unit cells  on both ends of the chain (as in the case with $\mathcal{I}$-symmetry). 
So the polarization per unit length for the  $t_1t_2t_2t_1$ chain with integer number of unit cells $N_{uc}$ and with the $N_{uc}$ lowest energy states filled is 
\begin{equation}
P/e=
\begin{cases}
0, \vert t_1\vert=\vert t_2\vert,\\
-\dfrac{1}{2}+\dfrac{N_{uc}-1}{N_{uc}}\dfrac{P_{CDAB}^{uc}}{a},\vert t_2\vert\ll \vert t_1\vert,\\
\dfrac{P_{ABCD}^{uc}}{a},\vert t_2\vert\gg\vert t_1\vert,
\end{cases}\end{equation}
where $P_{ABCD (CDAB)}^{uc}$ is the polarization of the unit cell $ABCD$ ($CDAB$) in the bottom (top) plot of Fig.~\ref{fig:cartoon}(b) and the $(-1/2)$ quantized term is due to the left edge state (ignoring finite size corrections)  of Fig.~\ref{fig:cartoon}(b). In Fig.~\ref{fig:cartoon}(c), we show these shifts in the quantization relation of the polarization in the case of 6 unit cells.
These arguments can be generalized for any choice of endings of the $t_1t_2t_2t_1$ chain.
Note that the charge density of the bottom band fixes the charge density of all bands (except at the edges) due to the chiral symmetry and the condition  that the total charge  gives one at every site.

To summarize, 1D topological insulators with non-centered
$\mathcal{I}$-axis in the unit cell for any choice of the unit cell show the following distinct features:
i) a $k$-dependent $\mathcal{I}$-operator within the unit
cell; 
ii) the need for a correction in the Zak's phase to recover $\pi$-quantization consistent with the eigenvalues of modified parity operators  at the $\mathcal{I}$-invariant momenta; 
iii) a sublattice chiral-like symmetry protecting finite energy edge states, reflecting the need, in contrast with what is usually assumed,  of explicit expressions for the respective symmetry operator not only in each $k$-subspace but also in the basis of the edge states;
iv) a shift of the  center of charge distribution of bulk bands  in relation to the center of the chain for  OBC commensurate with the unit cell.
%
These results can be straightforwardly generalized to quasi-1D models
(such as diamond chains \cite{Mukherjee2018,Kremer2018}) and ribbons with non-centered axes of $\mathcal{I}$-symmetry within the unit cell.


\section*{Acknowledgments}

\label{sec:acknowledments}

This work is funded by FEDER funds through the COMPETE 2020 Programme
and National Funds throught FCT - Portuguese Foundation for Science
and Technology under the project UID/CTM/50025/2013 and 
under the project PTDC/FIS-MAC/29291/2017. AMM acknowledges the financial support
from the FCT through the grant SFRH/PD/BD/108663/2015 and through the work contract CDL-CTTRI-147-ARH/2018, and from the Portuguese Institute for Nanostructures, Nanomodelling and Nanofabrication (I3N) through the grant BI/UI96/6376/2018. 
RGD appreciates the support by the Beijing CSRC.
We are grateful for useful discussions with E V Castro.


\bibliography{inversionsymmetry}

\begin{thebibliography}{29}%
\makeatletter
\providecommand \@ifxundefined [1]{%
 \@ifx{#1\undefined}
}%
\providecommand \@ifnum [1]{%
 \ifnum #1\expandafter \@firstoftwo
 \else \expandafter \@secondoftwo
 \fi
}%
\providecommand \@ifx [1]{%
 \ifx #1\expandafter \@firstoftwo
 \else \expandafter \@secondoftwo
 \fi
}%
\providecommand \natexlab [1]{#1}%
\providecommand \enquote  [1]{``#1''}%
\providecommand \bibnamefont  [1]{#1}%
\providecommand \bibfnamefont [1]{#1}%
\providecommand \citenamefont [1]{#1}%
\providecommand \href@noop [0]{\@secondoftwo}%
\providecommand \href [0]{\begingroup \@sanitize@url \@href}%
\providecommand \@href[1]{\@@startlink{#1}\@@href}%
\providecommand \@@href[1]{\endgroup#1\@@endlink}%
\providecommand \@sanitize@url [0]{\catcode `\\12\catcode `\$12\catcode
  `\&12\catcode `\#12\catcode `\^12\catcode `\_12\catcode `\%12\relax}%
\providecommand \@@startlink[1]{}%
\providecommand \@@endlink[0]{}%
\providecommand \url  [0]{\begingroup\@sanitize@url \@url }%
\providecommand \@url [1]{\endgroup\@href {#1}{\urlprefix }}%
\providecommand \urlprefix  [0]{URL }%
\providecommand \Eprint [0]{\href }%
\providecommand \doibase [0]{http://dx.doi.org/}%
\providecommand \selectlanguage [0]{\@gobble}%
\providecommand \bibinfo  [0]{\@secondoftwo}%
\providecommand \bibfield  [0]{\@secondoftwo}%
\providecommand \translation [1]{[#1]}%
\providecommand \BibitemOpen [0]{}%
\providecommand \bibitemStop [0]{}%
\providecommand \bibitemNoStop [0]{.\EOS\space}%
\providecommand \EOS [0]{\spacefactor3000\relax}%
\providecommand \BibitemShut  [1]{\csname bibitem#1\endcsname}%
\let\auto@bib@innerbib\@empty
\bibitem [{\citenamefont {Springborg}\ \emph {et~al.}(2004)\citenamefont
  {Springborg}, \citenamefont {Kirtman},\ and\ \citenamefont
  {Dong}}]{Springborg2004}%
  \BibitemOpen
  \bibfield  {author} {\bibinfo {author} {\bibfnamefont {M.}~\bibnamefont
  {Springborg}}, \bibinfo {author} {\bibfnamefont {B.}~\bibnamefont {Kirtman}},
  \ and\ \bibinfo {author} {\bibfnamefont {Y.}~\bibnamefont {Dong}},\ }\href
  {\doibase https://doi.org/10.1016/j.cplett.2004.08.067} {\bibfield  {journal}
  {\bibinfo  {journal} {Chem. Phys. Lett.}\ }\textbf {\bibinfo {volume}
  {396}},\ \bibinfo {pages} {404 } (\bibinfo {year} {2004})}\BibitemShut
  {NoStop}%
\bibitem [{\citenamefont {Kudin}\ \emph {et~al.}(2007)\citenamefont {Kudin},
  \citenamefont {Car},\ and\ \citenamefont {Resta}}]{Kudin2007}%
  \BibitemOpen
  \bibfield  {author} {\bibinfo {author} {\bibfnamefont {K.~N.}\ \bibnamefont
  {Kudin}}, \bibinfo {author} {\bibfnamefont {R.}~\bibnamefont {Car}}, \ and\
  \bibinfo {author} {\bibfnamefont {R.}~\bibnamefont {Resta}},\ }\href
  {\doibase 10.1063/1.2743018} {\bibfield  {journal} {\bibinfo  {journal} {J.
  Chem. Phys.}\ }\textbf {\bibinfo {volume} {126}},\ \bibinfo {pages} {234101}
  (\bibinfo {year} {2007})}\BibitemShut {NoStop}%
\bibitem [{\citenamefont {van Miert}\ \emph {et~al.}(2017)\citenamefont {van
  Miert}, \citenamefont {Ortix},\ and\ \citenamefont {Smith}}]{Miert2017}%
  \BibitemOpen
  \bibfield  {author} {\bibinfo {author} {\bibfnamefont {G.}~\bibnamefont {van
  Miert}}, \bibinfo {author} {\bibfnamefont {C.}~\bibnamefont {Ortix}}, \ and\
  \bibinfo {author} {\bibfnamefont {C.~M.}\ \bibnamefont {Smith}},\ }\href
  {http://stacks.iop.org/2053-1583/4/i=1/a=015023} {\bibfield  {journal}
  {\bibinfo  {journal} {2D Mater.}\ }\textbf {\bibinfo {volume} {4}},\ \bibinfo
  {pages} {015023} (\bibinfo {year} {2017})}\BibitemShut {NoStop}%
\bibitem [{\citenamefont {Rhim}\ \emph {et~al.}(2017)\citenamefont {Rhim},
  \citenamefont {Behrends},\ and\ \citenamefont {Bardarson}}]{Rhim2017}%
  \BibitemOpen
  \bibfield  {author} {\bibinfo {author} {\bibfnamefont {J.-W.}\ \bibnamefont
  {Rhim}}, \bibinfo {author} {\bibfnamefont {J.}~\bibnamefont {Behrends}}, \
  and\ \bibinfo {author} {\bibfnamefont {J.~H.}\ \bibnamefont {Bardarson}},\
  }\href {\doibase 10.1103/PhysRevB.95.035421} {\bibfield  {journal} {\bibinfo
  {journal} {Phys. Rev. B}\ }\textbf {\bibinfo {volume} {95}},\ \bibinfo
  {pages} {035421} (\bibinfo {year} {2017})}\BibitemShut {NoStop}%
\bibitem [{\citenamefont {Lin}\ and\ \citenamefont {Chou}(2018)}]{Lin2018}%
  \BibitemOpen
  \bibfield  {author} {\bibinfo {author} {\bibfnamefont {K.-S.}\ \bibnamefont
  {Lin}}\ and\ \bibinfo {author} {\bibfnamefont {M.-Y.}\ \bibnamefont {Chou}},\
  }\href {\doibase 10.1021/acs.nanolett.8b03417} {\bibfield  {journal}
  {\bibinfo  {journal} {Nano Lett.}\ }\textbf {\bibinfo {volume} {18}},\
  \bibinfo {pages} {7254} (\bibinfo {year} {2018})}\BibitemShut {NoStop}%
\bibitem [{\citenamefont {Arkinstall}\ \emph {et~al.}(2017)\citenamefont
  {Arkinstall}, \citenamefont {Teimourpour}, \citenamefont {Feng},
  \citenamefont {El-Ganainy},\ and\ \citenamefont
  {Schomerus}}]{Arkinstall2017}%
  \BibitemOpen
  \bibfield  {author} {\bibinfo {author} {\bibfnamefont {J.}~\bibnamefont
  {Arkinstall}}, \bibinfo {author} {\bibfnamefont {M.~H.}\ \bibnamefont
  {Teimourpour}}, \bibinfo {author} {\bibfnamefont {L.}~\bibnamefont {Feng}},
  \bibinfo {author} {\bibfnamefont {R.}~\bibnamefont {El-Ganainy}}, \ and\
  \bibinfo {author} {\bibfnamefont {H.}~\bibnamefont {Schomerus}},\ }\href
  {\doibase 10.1103/PhysRevB.95.165109} {\bibfield  {journal} {\bibinfo
  {journal} {Phys. Rev. B}\ }\textbf {\bibinfo {volume} {95}},\ \bibinfo
  {pages} {165109} (\bibinfo {year} {2017})}\BibitemShut {NoStop}%
\bibitem [{\citenamefont {Kremer}\ \emph {et~al.}(2018)\citenamefont {Kremer},
  \citenamefont {Petrides}, \citenamefont {Meyer}, \citenamefont {Heinrich},
  \citenamefont {Zilberberg},\ and\ \citenamefont {Szameit}}]{Kremer2018}%
  \BibitemOpen
  \bibfield  {author} {\bibinfo {author} {\bibfnamefont {M.}~\bibnamefont
  {Kremer}}, \bibinfo {author} {\bibfnamefont {I.}~\bibnamefont {Petrides}},
  \bibinfo {author} {\bibfnamefont {E.}~\bibnamefont {Meyer}}, \bibinfo
  {author} {\bibfnamefont {M.}~\bibnamefont {Heinrich}}, \bibinfo {author}
  {\bibfnamefont {O.}~\bibnamefont {Zilberberg}}, \ and\ \bibinfo {author}
  {\bibfnamefont {A.}~\bibnamefont {Szameit}},\ }\href@noop {} {\bibfield
  {journal} {\bibinfo  {journal} {arXiv e-prints}\ ,\ \bibinfo {pages}
  {arXiv:1805.05209}} (\bibinfo {year} {2018})}\BibitemShut {NoStop}%
\bibitem [{\citenamefont {Midya}\ \emph {et~al.}(2018)\citenamefont {Midya},
  \citenamefont {Walasik}, \citenamefont {Litchinitser},\ and\ \citenamefont
  {Feng}}]{Midya2018}%
  \BibitemOpen
  \bibfield  {author} {\bibinfo {author} {\bibfnamefont {B.}~\bibnamefont
  {Midya}}, \bibinfo {author} {\bibfnamefont {W.}~\bibnamefont {Walasik}},
  \bibinfo {author} {\bibfnamefont {N.~M.}\ \bibnamefont {Litchinitser}}, \
  and\ \bibinfo {author} {\bibfnamefont {L.}~\bibnamefont {Feng}},\ }\href
  {\doibase 10.1364/OL.43.004927} {\bibfield  {journal} {\bibinfo  {journal}
  {Opt. Lett.}\ }\textbf {\bibinfo {volume} {43}},\ \bibinfo {pages} {4927}
  (\bibinfo {year} {2018})}\BibitemShut {NoStop}%
\bibitem [{\citenamefont {Zhang}\ \emph {et~al.}(2019)\citenamefont {Zhang},
  \citenamefont {Teimourpour}, \citenamefont {Arkinstall}, \citenamefont {Pan},
  \citenamefont {Miao}, \citenamefont {Schomerus}, \citenamefont {El-Ganainy},\
  and\ \citenamefont {Feng}}]{Zhang2019}%
  \BibitemOpen
  \bibfield  {author} {\bibinfo {author} {\bibfnamefont {Z.}~\bibnamefont
  {Zhang}}, \bibinfo {author} {\bibfnamefont {M.~H.}\ \bibnamefont
  {Teimourpour}}, \bibinfo {author} {\bibfnamefont {J.}~\bibnamefont
  {Arkinstall}}, \bibinfo {author} {\bibfnamefont {M.}~\bibnamefont {Pan}},
  \bibinfo {author} {\bibfnamefont {P.}~\bibnamefont {Miao}}, \bibinfo {author}
  {\bibfnamefont {H.}~\bibnamefont {Schomerus}}, \bibinfo {author}
  {\bibfnamefont {R.}~\bibnamefont {El-Ganainy}}, \ and\ \bibinfo {author}
  {\bibfnamefont {L.}~\bibnamefont {Feng}},\ }\href {\doibase
  10.1002/lpor.201800202} {\bibfield  {journal} {\bibinfo  {journal} {Laser
  Photonics Rev.}\ }\textbf {\bibinfo {volume} {13}},\ \bibinfo {pages}
  {1800202} (\bibinfo {year} {2019})}\BibitemShut {NoStop}%
\bibitem [{\citenamefont {Pelegr\'{\i}}\ \emph
  {et~al.}(2019{\natexlab{a}})\citenamefont {Pelegr\'{\i}}, \citenamefont
  {Marques}, \citenamefont {Dias}, \citenamefont {Daley}, \citenamefont
  {Ahufinger},\ and\ \citenamefont {Mompart}}]{Pelegri2019}%
  \BibitemOpen
  \bibfield  {author} {\bibinfo {author} {\bibfnamefont {G.}~\bibnamefont
  {Pelegr\'{\i}}}, \bibinfo {author} {\bibfnamefont {A.~M.}\ \bibnamefont
  {Marques}}, \bibinfo {author} {\bibfnamefont {R.~G.}\ \bibnamefont {Dias}},
  \bibinfo {author} {\bibfnamefont {A.~J.}\ \bibnamefont {Daley}}, \bibinfo
  {author} {\bibfnamefont {V.}~\bibnamefont {Ahufinger}}, \ and\ \bibinfo
  {author} {\bibfnamefont {J.}~\bibnamefont {Mompart}},\ }\href {\doibase
  10.1103/PhysRevA.99.023612} {\bibfield  {journal} {\bibinfo  {journal} {Phys.
  Rev. A}\ }\textbf {\bibinfo {volume} {99}},\ \bibinfo {pages} {023612}
  (\bibinfo {year} {2019}{\natexlab{a}})}\BibitemShut {NoStop}%
\bibitem [{\citenamefont {Pelegr\'{\i}}\ \emph
  {et~al.}(2019{\natexlab{b}})\citenamefont {Pelegr\'{\i}}, \citenamefont
  {Marques}, \citenamefont {Dias}, \citenamefont {Daley}, \citenamefont
  {Mompart},\ and\ \citenamefont {Ahufinger}}]{Pelegri2019b}%
  \BibitemOpen
  \bibfield  {author} {\bibinfo {author} {\bibfnamefont {G.}~\bibnamefont
  {Pelegr\'{\i}}}, \bibinfo {author} {\bibfnamefont {A.~M.}\ \bibnamefont
  {Marques}}, \bibinfo {author} {\bibfnamefont {R.~G.}\ \bibnamefont {Dias}},
  \bibinfo {author} {\bibfnamefont {A.~J.}\ \bibnamefont {Daley}}, \bibinfo
  {author} {\bibfnamefont {J.}~\bibnamefont {Mompart}}, \ and\ \bibinfo
  {author} {\bibfnamefont {V.}~\bibnamefont {Ahufinger}},\ }\href {\doibase
  10.1103/PhysRevA.99.023613} {\bibfield  {journal} {\bibinfo  {journal} {Phys.
  Rev. A}\ }\textbf {\bibinfo {volume} {99}},\ \bibinfo {pages} {023613}
  (\bibinfo {year} {2019}{\natexlab{b}})}\BibitemShut {NoStop}%
\bibitem [{\citenamefont {Mei}\ \emph {et~al.}(2012)\citenamefont {Mei},
  \citenamefont {Zhu}, \citenamefont {Zhang}, \citenamefont {Oh},\ and\
  \citenamefont {Goldman}}]{Mei2012}%
  \BibitemOpen
  \bibfield  {author} {\bibinfo {author} {\bibfnamefont {F.}~\bibnamefont
  {Mei}}, \bibinfo {author} {\bibfnamefont {S.-L.}\ \bibnamefont {Zhu}},
  \bibinfo {author} {\bibfnamefont {Z.-M.}\ \bibnamefont {Zhang}}, \bibinfo
  {author} {\bibfnamefont {C.~H.}\ \bibnamefont {Oh}}, \ and\ \bibinfo {author}
  {\bibfnamefont {N.}~\bibnamefont {Goldman}},\ }\href {\doibase
  10.1103/PhysRevA.85.013638} {\bibfield  {journal} {\bibinfo  {journal} {Phys.
  Rev. A}\ }\textbf {\bibinfo {volume} {85}},\ \bibinfo {pages} {013638}
  (\bibinfo {year} {2012})}\BibitemShut {NoStop}%
\bibitem [{\citenamefont {Lang}\ \emph {et~al.}(2012)\citenamefont {Lang},
  \citenamefont {Cai},\ and\ \citenamefont {Chen}}]{Lang2012}%
  \BibitemOpen
  \bibfield  {author} {\bibinfo {author} {\bibfnamefont {L.-J.}\ \bibnamefont
  {Lang}}, \bibinfo {author} {\bibfnamefont {X.}~\bibnamefont {Cai}}, \ and\
  \bibinfo {author} {\bibfnamefont {S.}~\bibnamefont {Chen}},\ }\href {\doibase
  10.1103/PhysRevLett.108.220401} {\bibfield  {journal} {\bibinfo  {journal}
  {Phys. Rev. Lett.}\ }\textbf {\bibinfo {volume} {108}},\ \bibinfo {pages}
  {220401} (\bibinfo {year} {2012})}\BibitemShut {NoStop}%
\bibitem [{\citenamefont {Zhu}\ \emph {et~al.}(2013)\citenamefont {Zhu},
  \citenamefont {Wang}, \citenamefont {Chan},\ and\ \citenamefont
  {Duan}}]{Zhu2013}%
  \BibitemOpen
  \bibfield  {author} {\bibinfo {author} {\bibfnamefont {S.-L.}\ \bibnamefont
  {Zhu}}, \bibinfo {author} {\bibfnamefont {Z.-D.}\ \bibnamefont {Wang}},
  \bibinfo {author} {\bibfnamefont {Y.-H.}\ \bibnamefont {Chan}}, \ and\
  \bibinfo {author} {\bibfnamefont {L.-M.}\ \bibnamefont {Duan}},\ }\href
  {\doibase 10.1103/PhysRevLett.110.075303} {\bibfield  {journal} {\bibinfo
  {journal} {Phys. Rev. Lett.}\ }\textbf {\bibinfo {volume} {110}},\ \bibinfo
  {pages} {075303} (\bibinfo {year} {2013})}\BibitemShut {NoStop}%
\bibitem [{\citenamefont {Qin}\ \emph {et~al.}(2017)\citenamefont {Qin},
  \citenamefont {Mei}, \citenamefont {Ke}, \citenamefont {Zhang},\ and\
  \citenamefont {Lee}}]{Qin2017}%
  \BibitemOpen
  \bibfield  {author} {\bibinfo {author} {\bibfnamefont {X.}~\bibnamefont
  {Qin}}, \bibinfo {author} {\bibfnamefont {F.}~\bibnamefont {Mei}}, \bibinfo
  {author} {\bibfnamefont {Y.}~\bibnamefont {Ke}}, \bibinfo {author}
  {\bibfnamefont {L.}~\bibnamefont {Zhang}}, \ and\ \bibinfo {author}
  {\bibfnamefont {C.}~\bibnamefont {Lee}},\ }\href {\doibase
  10.1103/PhysRevB.96.195134} {\bibfield  {journal} {\bibinfo  {journal} {Phys.
  Rev. B}\ }\textbf {\bibinfo {volume} {96}},\ \bibinfo {pages} {195134}
  (\bibinfo {year} {2017})}\BibitemShut {NoStop}%
\bibitem [{\citenamefont {Martinez~Alvarez}\ and\ \citenamefont
  {Coutinho-Filho}(2019)}]{Alvarez2019}%
  \BibitemOpen
  \bibfield  {author} {\bibinfo {author} {\bibfnamefont {V.~M.}\ \bibnamefont
  {Martinez~Alvarez}}\ and\ \bibinfo {author} {\bibfnamefont {M.~D.}\
  \bibnamefont {Coutinho-Filho}},\ }\href {\doibase 10.1103/PhysRevA.99.013833}
  {\bibfield  {journal} {\bibinfo  {journal} {Phys. Rev. A}\ }\textbf {\bibinfo
  {volume} {99}},\ \bibinfo {pages} {013833} (\bibinfo {year}
  {2019})}\BibitemShut {NoStop}%
\bibitem [{\citenamefont {Alexandradinata}\ \emph {et~al.}(2014)\citenamefont
  {Alexandradinata}, \citenamefont {Dai},\ and\ \citenamefont
  {Bernevig}}]{Alexandradinata2014}%
  \BibitemOpen
  \bibfield  {author} {\bibinfo {author} {\bibfnamefont {A.}~\bibnamefont
  {Alexandradinata}}, \bibinfo {author} {\bibfnamefont {X.}~\bibnamefont
  {Dai}}, \ and\ \bibinfo {author} {\bibfnamefont {B.~A.}\ \bibnamefont
  {Bernevig}},\ }\href {\doibase 10.1103/PhysRevB.89.155114} {\bibfield
  {journal} {\bibinfo  {journal} {Phys. Rev. B}\ }\textbf {\bibinfo {volume}
  {89}},\ \bibinfo {pages} {155114} (\bibinfo {year} {2014})}\BibitemShut
  {NoStop}%
\bibitem [{\citenamefont {Asb\'oth}\ \emph {et~al.}(2016)\citenamefont
  {Asb\'oth}, \citenamefont {Oroszl\'any},\ and\ \citenamefont
  {P\'alyi}}]{Asboth2016}%
  \BibitemOpen
  \bibfield  {author} {\bibinfo {author} {\bibfnamefont {J.~K.}\ \bibnamefont
  {Asb\'oth}}, \bibinfo {author} {\bibfnamefont {L.}~\bibnamefont
  {Oroszl\'any}}, \ and\ \bibinfo {author} {\bibfnamefont {A.}~\bibnamefont
  {P\'alyi}},\ }\href@noop {} {\emph {\bibinfo {title} {A Short Course on
  Topological Insulators}}}\ (\bibinfo  {publisher} {Springer, Berlin},\
  \bibinfo {year} {2016})\BibitemShut {NoStop}%
\bibitem [{\citenamefont {Zak}(1989)}]{Zak1989}%
  \BibitemOpen
  \bibfield  {author} {\bibinfo {author} {\bibfnamefont {J.}~\bibnamefont
  {Zak}},\ }\href {\doibase 10.1103/PhysRevLett.62.2747} {\bibfield  {journal}
  {\bibinfo  {journal} {Phys. Rev. Lett.}\ }\textbf {\bibinfo {volume} {62}},\
  \bibinfo {pages} {2747} (\bibinfo {year} {1989})}\BibitemShut {NoStop}%
\bibitem [{\citenamefont {Eliashvili}\ \emph {et~al.}(2017)\citenamefont
  {Eliashvili}, \citenamefont {Kereselidze}, \citenamefont {Tsitsishvili},\
  and\ \citenamefont {Tsitsishvili}}]{Eliashvili2017}%
  \BibitemOpen
  \bibfield  {author} {\bibinfo {author} {\bibfnamefont {M.}~\bibnamefont
  {Eliashvili}}, \bibinfo {author} {\bibfnamefont {D.}~\bibnamefont
  {Kereselidze}}, \bibinfo {author} {\bibfnamefont {G.}~\bibnamefont
  {Tsitsishvili}}, \ and\ \bibinfo {author} {\bibfnamefont {M.}~\bibnamefont
  {Tsitsishvili}},\ }\href {\doibase 10.7566/JPSJ.86.074712} {\bibfield
  {journal} {\bibinfo  {journal} {J. Phys. Soc. Jpn.}\ }\textbf {\bibinfo
  {volume} {86}},\ \bibinfo {pages} {074712} (\bibinfo {year}
  {2017})}\BibitemShut {NoStop}%
\bibitem [{\citenamefont {Maffei}\ \emph {et~al.}(2018)\citenamefont {Maffei},
  \citenamefont {Dauphin}, \citenamefont {Cardano}, \citenamefont
  {Lewenstein},\ and\ \citenamefont {Massignan}}]{Maffei2018}%
  \BibitemOpen
  \bibfield  {author} {\bibinfo {author} {\bibfnamefont {M.}~\bibnamefont
  {Maffei}}, \bibinfo {author} {\bibfnamefont {A.}~\bibnamefont {Dauphin}},
  \bibinfo {author} {\bibfnamefont {F.}~\bibnamefont {Cardano}}, \bibinfo
  {author} {\bibfnamefont {M.}~\bibnamefont {Lewenstein}}, \ and\ \bibinfo
  {author} {\bibfnamefont {P.}~\bibnamefont {Massignan}},\ }\href
  {http://stacks.iop.org/1367-2630/20/i=1/a=013023} {\bibfield  {journal}
  {\bibinfo  {journal} {New J. Phys.}\ }\textbf {\bibinfo {volume} {20}},\
  \bibinfo {pages} {013023} (\bibinfo {year} {2018})}\BibitemShut {NoStop}%
\bibitem [{\citenamefont {Ryu}\ and\ \citenamefont {Hatsugai}(2002)}]{Ryu2002}%
  \BibitemOpen
  \bibfield  {author} {\bibinfo {author} {\bibfnamefont {S.}~\bibnamefont
  {Ryu}}\ and\ \bibinfo {author} {\bibfnamefont {Y.}~\bibnamefont {Hatsugai}},\
  }\href {\doibase 10.1103/PhysRevLett.89.077002} {\bibfield  {journal}
  {\bibinfo  {journal} {Phys. Rev. Lett.}\ }\textbf {\bibinfo {volume} {89}},\
  \bibinfo {pages} {077002} (\bibinfo {year} {2002})}\BibitemShut {NoStop}%
\bibitem [{\citenamefont {Delplace}\ \emph {et~al.}(2011)\citenamefont
  {Delplace}, \citenamefont {Ullmo},\ and\ \citenamefont
  {Montambaux}}]{Delplace2011}%
  \BibitemOpen
  \bibfield  {author} {\bibinfo {author} {\bibfnamefont {P.}~\bibnamefont
  {Delplace}}, \bibinfo {author} {\bibfnamefont {D.}~\bibnamefont {Ullmo}}, \
  and\ \bibinfo {author} {\bibfnamefont {G.}~\bibnamefont {Montambaux}},\
  }\href {\doibase 10.1103/PhysRevB.84.195452} {\bibfield  {journal} {\bibinfo
  {journal} {Phys. Rev. B}\ }\textbf {\bibinfo {volume} {84}},\ \bibinfo
  {pages} {195452} (\bibinfo {year} {2011})}\BibitemShut {NoStop}%
\bibitem [{\citenamefont {Banchi}\ and\ \citenamefont
  {Vaia}(2013)}]{Banchi2013}%
  \BibitemOpen
  \bibfield  {author} {\bibinfo {author} {\bibfnamefont {L.}~\bibnamefont
  {Banchi}}\ and\ \bibinfo {author} {\bibfnamefont {R.}~\bibnamefont {Vaia}},\
  }\href {\doibase 10.1063/1.4797477} {\bibfield  {journal} {\bibinfo
  {journal} {J. Math. Phys.}\ }\textbf {\bibinfo {volume} {54}},\ \bibinfo
  {pages} {043501} (\bibinfo {year} {2013})}\BibitemShut {NoStop}%
\bibitem [{\citenamefont {H\"ugel}\ and\ \citenamefont
  {Paredes}(2014)}]{Hugel2014}%
  \BibitemOpen
  \bibfield  {author} {\bibinfo {author} {\bibfnamefont {D.}~\bibnamefont
  {H\"ugel}}\ and\ \bibinfo {author} {\bibfnamefont {B.}~\bibnamefont
  {Paredes}},\ }\href {\doibase 10.1103/PhysRevA.89.023619} {\bibfield
  {journal} {\bibinfo  {journal} {Phys. Rev. A}\ }\textbf {\bibinfo {volume}
  {89}},\ \bibinfo {pages} {023619} (\bibinfo {year} {2014})}\BibitemShut
  {NoStop}%
\bibitem [{\citenamefont {Duncan}\ \emph {et~al.}(2018)\citenamefont {Duncan},
  \citenamefont {\"Ohberg},\ and\ \citenamefont {Valiente}}]{Duncan2018}%
  \BibitemOpen
  \bibfield  {author} {\bibinfo {author} {\bibfnamefont {C.~W.}\ \bibnamefont
  {Duncan}}, \bibinfo {author} {\bibfnamefont {P.}~\bibnamefont {\"Ohberg}}, \
  and\ \bibinfo {author} {\bibfnamefont {M.}~\bibnamefont {Valiente}},\ }\href
  {\doibase 10.1103/PhysRevB.97.195439} {\bibfield  {journal} {\bibinfo
  {journal} {Phys. Rev. B}\ }\textbf {\bibinfo {volume} {97}},\ \bibinfo
  {pages} {195439} (\bibinfo {year} {2018})}\BibitemShut {NoStop}%
\bibitem [{Note1()}]{Note1}%
  \BibitemOpen
  \bibinfo {note} {See Supplemental Material for additional details on deriving
  the edge states from the general form of (\ref {eq:eigenedge})}\BibitemShut
  {NoStop}%
\bibitem [{Note2()}]{Note2}%
  \BibitemOpen
  \bibinfo {note} {See Supplemental Material for additional details on the
  derivation of $\protect \mathaccentV
  {hat}05E{C}_{1/2}^{edge}(c)$}\BibitemShut {NoStop}%
\bibitem [{\citenamefont {Mukherjee}\ \emph {et~al.}(2018)\citenamefont
  {Mukherjee}, \citenamefont {Di~Liberto}, \citenamefont {\"Ohberg},
  \citenamefont {Thomson},\ and\ \citenamefont {Goldman}}]{Mukherjee2018}%
  \BibitemOpen
  \bibfield  {author} {\bibinfo {author} {\bibfnamefont {S.}~\bibnamefont
  {Mukherjee}}, \bibinfo {author} {\bibfnamefont {M.}~\bibnamefont
  {Di~Liberto}}, \bibinfo {author} {\bibfnamefont {P.}~\bibnamefont
  {\"Ohberg}}, \bibinfo {author} {\bibfnamefont {R.~R.}\ \bibnamefont
  {Thomson}}, \ and\ \bibinfo {author} {\bibfnamefont {N.}~\bibnamefont
  {Goldman}},\ }\href {\doibase 10.1103/PhysRevLett.121.075502} {\bibfield
  {journal} {\bibinfo  {journal} {Phys. Rev. Lett.}\ }\textbf {\bibinfo
  {volume} {121}},\ \bibinfo {pages} {075502} (\bibinfo {year}
  {2018})}\BibitemShut {NoStop}%
\end{thebibliography}%

\pagebreak
\section*{Supplemental material}
\section{Edge state solutions of Eq.~14}

Let us address edge-like states in an infinite $t_{2}t_{1}t_{1}t_{2}$
chain. As mentioned in the main text, these states
are not normalizable, but if they have zeros of amplitude at certain
sites of the chain, open boundary conditions  (OBC) may be introduced at these sites and one obtains
an eigenstate of the finite chain which is orthogonal to the ``harmonic''
eigenstates. So we look for solutions of the type
\begin{equation}
c^{j}\begin{bmatrix}\psi_{A}\\
\psi_{B}\\
\psi_{C}\\
\psi_{D}
\end{bmatrix}.
\end{equation}

We have four equations from the eigenvalue relation
\begin{eqnarray}
\varepsilon c^{j}\psi_{A} & = & t_{2}\psi_{B}c^{j}+t_{2}\psi_{D}c^{j-1},\\
\varepsilon c^{j}\psi_{B} & = & t_{2}\psi_{A}c^{j}+t_{1}\psi_{C}c^{j},\\
\varepsilon c^{j}\psi_{C} & = & t_{1}\psi_{B}c^{j}+t_{1}\psi_{D}c^{j},\\
\varepsilon c^{j}\psi_{D} & = & t_{1}\psi_{C}c^{j}+t_{2}\psi_{A}c^{j+1},
\end{eqnarray}
which can be rewritten as a matrix equation
\begin{equation}
\varepsilon\begin{bmatrix}\psi_{A}\\
\psi_{B}\\
\psi_{C}\\
\psi_{D}
\end{bmatrix}=\begin{bmatrix}0 & t_{2} & 0 & t_{2}/c\\
t_{2} & 0 & t_{1} & 0\\
0 & t_{1} & 0 & t_{1}\\
t_{2}c & 0 & t_{1} & 0
\end{bmatrix}\begin{bmatrix}\psi_{A}\\
\psi_{B}\\
\psi_{C}\\
\psi_{D}
\end{bmatrix},
\end{equation}
leading to four energies (for $t_{2}=1$ and $t_1=t$)
\begin{equation}
\varepsilon=\pm\frac{\sqrt{c^{2}\left(t^{2}+1\right)\pm\sqrt{c^{3}\left(c+t^{2}\right)\left(ct^{2}+1\right)}}}{c},
\end{equation}
The respective eigenstates
are 
\begin{equation}
\begin{bmatrix}\varepsilon(\varepsilon^{2}-2t^{2})\\
(\varepsilon^{2}-t^{2}+ct^{2})\\
(1+c)t\varepsilon\\
(c\varepsilon^{2}+t^{2}-ct^{2})
\end{bmatrix}.
\end{equation}

We now require that one of the amplitudes is zero so that this state
is an eigenstate of the finite chain with OBC.  If we choose the
first component, this leads to $\varepsilon=\pm\sqrt{2}t$ and $c=-1$
($k=\pi$ state) or to $\varepsilon=0$ and $c=1$ ($k=0$ state).
This implies that no edge state will be present at a left edge that
ends with a B site or at a right edge that ends with a D site. 

If we choose the second component, besides solutions that are not
edge states (a $k=0$ state with zero energy), we have 
\begin{eqnarray}
\varepsilon & = & \pm\sqrt{1+t^{2}},\\
c & = & -1/t^{2},
\end{eqnarray}
and the eigenstate is 
\begin{equation}
\begin{bmatrix}\pm\sqrt{1+t^{2}}(1-t^{2})\\
0\\
\mp\sqrt{1+t^{2}}(1-t^{2})/t\\
-1/t^{2}+t^{2}
\end{bmatrix}
\end{equation}
Recalling that we are working with 
non-normalized edge states, we may divide the previous state by the first component leading to 
\begin{equation}
\begin{bmatrix}1\\
0\\
-1/t\\
-\varepsilon/t^2
\end{bmatrix}
\xrightarrow[]{\times (-t^2)}
\begin{bmatrix}-t^2\\
0\\
t\\
\varepsilon
\end{bmatrix}.
\end{equation} 
which is the form of (ii)$\ket{u(\varepsilon,c=-\frac{1}{t^{2}})}=(-t^{2},0,t,\varepsilon)^{T}$ shown in the main text below Eq.~14.
A similar procedure is followed in order to obtain $\ket{u(\varepsilon,c=-t^{2})}=(1,\varepsilon,t,0)^{T}$.

\section{How to derive Eq.~15 from Eq.~13}

The   unitary Hermitian operator
$\hat{C}_{1/2}=\sum\limits _{k}\hat{C}_{1/2}(k)$,
with
\begin{eqnarray}
\hat{C}_{1/2}(k) & = & \ket{u_{1}(k)}\bra{u_{2}(k)}+\ket{u_{3}(k)}\bra{u_{4}(k)}+H.c.,
\end{eqnarray}
in the basis ${\vert k,A\rangle,\vert k,C\rangle,\vert k,B\rangle,\vert k,D\rangle}$  [where $A,B,C,D$ are the sites of  the unit cell of the $t_{1}t_{1}t_{2}t_{2}$ chain shown in Fig.~1(b) of the main text]
is given by
\begin{equation}
\hat{C}_{1/2}(k)=\left[\begin{array}{cc}
\text{sgn}(k) [\sin(\frac{k}{2})\sigma_{x}-\cos(\frac{k}{2})\sigma_{y} ]& 0\\
0 & \sigma_{z}
\end{array}\right],
\end{equation}
where $\sigma_{\alpha}$ are the Pauli matrices:
\begin{align}
\sigma_x &=
\begin{bmatrix}
0&1\\
1&0
\end{bmatrix}, \\
\sigma_y &=
\begin{bmatrix}
0&-i\\
i&0
\end{bmatrix}, \\
\sigma_z &=
\begin{bmatrix}
1&0\\
0&-1
\end{bmatrix} \,.
\end{align}

The  explicit form of the  unitary Hermitian operator
$\hat{C}_{1/2}(k)$ substituting  the Pauli matrices for $k>0$
is
\begin{eqnarray}
\hat{C}_{1/2}(k)& = & \left[\begin{array}{cccc}
0 & ie^{ik/2} & 0 & 0\\
-ie^{-ik/2} & 0 & 0 & 0\\
0 & 0 & 1 & 0\\
0 & 0 & 0 & -1
\end{array}\right].
\end{eqnarray}
Since the
$\hat{C}_{1/2}$ operator has a $k$-dependent $\hat{C}_{1/2}(k)$
matrix representation, the matrix representation in the edge states subspace
of the $\hat{C}_{1/2}$ operator is obtained by analytical continuation,
that is, $\hat{C}_{1/2}^{edge}(c)=\hat{C}_{1/2}^{Bloch}(e^{ik}\rightarrow c)$ with $c=-e^{-\alpha}=e^{i(\pi+i\alpha)}$ ,
and in the $\{\ket{A},\ket{C},\ket{B},\ket{D}\}$ basis of each unit cell, it is given by

\begin{equation}
\hat{C}_{1/2}^{edge}(c)=\begin{bmatrix}0 & \frac{1}{\sqrt{\vert c\vert}} & 0 & 0\\
\sqrt{\vert c\vert} & 0 & 0 & 0\\
0 & 0 & 1 & 0\\
0 & 0 & 0 & -1
\end{bmatrix}.
\end{equation}
If one chooses $c=-e^{-\alpha}=e^{i(-\pi+i\alpha)}$  and uses the matrix representation of $\hat{C}_{1/2}$ operator with $k<0$, one obtains the same result. 

\end{document}